\documentclass[aps,showpacs,superscriptaddress,preprint]{revtex4-2}

\usepackage{titlesec}
\usepackage{amssymb,amsmath,amsfonts,latexsym,graphicx,epsfig,bm}
\usepackage{epstopdf}

\usepackage{tabularx} 
\usepackage{multirow}

\usepackage{enumitem} 

\titleformat{\section}{\large\bfseries}{\thesection}{1em}{}

\newcommand{\bea}{\begin{eqnarray}}
	\newcommand{\ena}{\end{eqnarray}}
\newcommand{\nn}{\nonumber\\}
\newcommand{\be}{\begin{equation}}
	\newcommand{\en}{\end{equation}}

\newcommand{\ed}{\end{document}}

\newcommand{\slp}{p\kern-5pt/}

\makeatletter
\def\switch@array{}  
\makeatother

\begin{document}

\title{Radiative and Dalitz decays of \boldmath{$\Upsilon(1S)$} \\in the light of the ATOMKI \boldmath{$X17$} anomaly}

\author{C.~T.~Tran}
\email{thangtc@hcmute.edu.vn}
\affiliation{
	\hbox{Department of Engineering Physics, HCMC University of Technology and Engineering, }\\
	Vo Van Ngan 1, 700000 Ho Chi Minh City, Vietnam}

\author{M. A. Ivanov}
\affiliation{Bogoliubov Laboratory of Theoretical Physics, 
	Joint Institute for Nuclear Research, 141980 Dubna, Russia}


\begin{abstract}
	We study the radiative decay $\Upsilon(1S)\to \eta_b\gamma$ and the Dalitz decay $\Upsilon(1S)\to \eta_b e^+ e^-$ within the framework of the Covariant Confined Quark Model. We provide theoretical predictions for the hadronic form factor, the radiative decay constant, and the decay branching fractions. Within the Standard Model we predict $\Gamma(\Upsilon(1S)\to\eta_b \gamma) = 9.3(9)\, \textrm{eV}$ and $\Gamma(\Upsilon(1S)\to\eta_b e^+ e^-) = 4.8(5)\times 10^{-2}\, \textrm{eV}$. The contribution of the hypothetical ATOMKI $X17$ vector boson to the Dalitz channel is also investigated.  
\end{abstract}

\maketitle
\newpage

\section{Introduction}
The ATOMKI Collaboration's observation of a persistent 17~MeV anomaly in $^8\text{Be}$~\cite{Krasznahorkay:2015iga, Krasznahorkay:2019lgi}, $^4\text{He}$~\cite{ Krasznahorkay:2021joi}, and $^{12}\text{C}$~\cite{Krasznahorkay:2022pxs} nuclear transitions has sparked significant interest, suggesting a new light boson $X17$ beyond the Standard Model (SM). While quantum electrodynamics predicts a smooth, monotonic decline in the internal pair creation correlation at large opening angles~\cite{Rose:1949zz, Schluter:1981cjo}, experimental data across multiple nuclear reactions consistently reveals a ``bump" signature at approximately 17~MeV. Independent experimental efforts have sought to verify these findings. While the NA64~\cite{NA64:2018lsq, NA64:2019auh, NA64:2020xxh} and MEG~II~\cite{MEGII:2024urz} Collaborations have set stringent coupling limits, recent results from VNU~\cite{Anh:2024req} and the PADME experiment~\cite{PADME:2025dla, PADME:2025dvz} (which reported a 2.5$\sigma$ excess at 16.90~MeV) provide further motivation for a New Physics (NP) interpretation. For a review of the $X17$ anomaly, see Ref.~\cite{Alves:2023ree}.

Theoretically, the ATOMKI anomaly has been interpreted through various scenarios, including protophobic vector~\cite{Feng:2016jff, Feng:2016ysn, Zhang:2020ukq,Pulice:2019xel}, axial-vector~\cite{Kozaczuk:2016nma, Barducci:2022lqd,Seto:2020jal,Nomura:2020kcw}, and pseudoscalar models~\cite{Ellwanger:2016wfe}, often in connection with other unresolved issues such as the $(g-2)_\mu$ anomaly and the proton charge radius puzzle~\cite{Kirpichnikov:2020tcf, DiLuzio:2025ojt, Barducci:2025hpg}. Among these, the vector boson hypothesis is of great interest since it opens the door to probing the fifth fundamental force~\cite{Feng:2020mbt}. This hypothesis has extended the search into the particle physics sector, particularly through Dalitz decays ($V \to P e^+ e^-$) of vector mesons~\cite{Fu:2011yy, Castro:2021gdf, Ban:2020uii, Lee:2025lwv,Colangelo:2025yud}. Notably, the BESIII collaboration recently observed a 3.5$\sigma$ excess~\cite{BESIII:2021vyq} in $D^{*0} \to D^0 e^+ e^-$ relative to Vector Meson Dominance (VMD) predictions. While several studies have utilized VMD and Heavy Quark Effective Theory (HQET) to model these transitions in the charm and beauty sectors, conflicting results and the inherent limitations of these models highlight the need for more rigorous hadronic form factor calculations to definitively probe for $X17$ signals. 

In a recent paper~\cite{Tran:2025fhb}, we investigated the effects of the hypothetical vector boson $X17$ on the Dalitz decays $D^*_{(s)}\to D_{(s)}e^+e^-$, $B^*_{(s)}\to B_{(s)}e^+e^-$, and $J/\psi\to \eta_ce^+e^-$ using hadronic form factors calculated within the framework of the Covariant Confined Quark Model (CCQM), without relying on HQET or the VMD model. While our SM predictions were consistent with existing VMD-based literature, the calculated NP contributions from $X17$ revealed discrepancies of up to an order of magnitude between the two theoretical approaches. Besides, among the seven channels examined, $D^{*0} \to D^0 e^+ e^-$ and $D^{*+}_s \to D^+_s e^+ e^-$ were identified as the most sensitive to $X17$ effects. Interestingly, while the $D_s$ channel data from CLEO~\cite{CLEO:2011mla} aligns well with our predictions when $X17$ is included, the $3.5\sigma$ excess observed by BESIII in $D^{*0} \to D^0 e^+ e^-$~\cite{BESIII:2021vyq} remains unexplained, as both CCQM and VMD suggest an $X17$ contribution far too small to account for the deviation.

This tension in the charm sector underscores the necessity for independent theoretical explorations and broader experimental searches. In particular, the heavy quarkonium sector offers a higher mass scale and a different kinematic regime to test the consistency of the $X17$ hypothesis. 
Consequently, this study extends the CCQM analysis to the bottomonium sector, specifically targeting the $\Upsilon(1S) \to \eta_b e^+ e^-$ transition. By investigating this new channel, we aim to determine if the sensitivities observed in charm and light beauty mesons persist in $b\bar{b}$ systems, potentially providing the missing pieces to the current $X17$ puzzle. As a by-product, we also calculate the corresponding radiative decay $\Upsilon(1S) \to \eta_b \gamma$ and compare with other theoretical calculations.

The paper is organized as follows. In Sec.~\ref{sec:formalism} we present the relevant theoretical formalism for the calculation of the radiative and Dalitz decays of $\Upsilon(1S)$. The section includes a brief description of the CCQM. In Sec.~\ref{sec:result} we provide the numerical results for the form factors, the radiative decay constant, and the branching fractions. A comparison with the literature is also provided. Finally, a brief summary is given in Sec.~\ref{sec:sum}.

\section{Formalism}
\label{sec:formalism}
The quark-photon coupling is described by the interaction Lagrangian
\begin{equation}
	\mathcal{L}^{\mathrm{int}}_{\mathrm{em}}(x) = eA_\mu(x)J_{\mathrm{em}}^\mu(x),\qquad J_{\mathrm{em}}^\mu(x) = e_b\bar{b}(x)\gamma^\mu b(x),
\end{equation}
where $e_b$ is the charge of $b$ quark in units of $e$. Similarly, for the interaction between the quarks and the $X17$ boson one has
\begin{equation}
	\mathcal{L}^{\mathrm{int}}_{X}(x) = eX_\mu(x)J_{X}^\mu(x),\qquad J_{X}^\mu(x) = \varepsilon_b\bar{b}(x)\gamma^\mu b(x),
\end{equation}
where $\varepsilon_b$ is the coupling constant between the $X17$ boson and $b$ quark.

The invariant matrix elements of the Dalitz decay $\Upsilon(p,\epsilon_\Upsilon)\to \eta_b(p^\prime)e^+(q_+)e^-(q_-)$ are written as
\begin{eqnarray}
	\label{eq:M}
	\mathcal{M} &=& \mathcal{M}^\gamma + \mathcal{M}^X,\nn
	\mathcal{M}^\gamma &=& T_\mu^\gamma \frac{g^{\mu\nu}}{q^2+i\epsilon} (ie)\bar{u}(q_-)\gamma_\nu v(q_+),\nn
	\mathcal{M}^X &=& T_\mu^X \frac{g^{\mu\nu}-q^\mu q^\nu/m_X^2}{q^2-m_X^2+i m_X\Gamma_X} (ie\varepsilon_e)\bar{u}(q_-)\gamma_\nu v(q_+),
\end{eqnarray}
where $e\varepsilon_e$ is the coupling constant between the $X17$ boson and electron (positron), $\epsilon_\Upsilon$ is the polarization vector of $\Upsilon(1S)$, and $T_\mu^{\gamma,X}$ are the hadronic amplitudes of the transitions $\Upsilon(1S)\to \eta_b\gamma$ and $\Upsilon(1S)\to \eta_b X$, respectively. The hadronic transition amplitudes are parametrized in terms of the transition form factors $G_{\gamma}(q^2)$ and $G_{X}(q^2)$ as follows:
\begin{eqnarray}
	\label{eq:T}
	T_\mu^\gamma &\equiv& \left\langle \eta_b(p^\prime)|J_\mu^{\textrm{em}}|\Upsilon(p,\epsilon_\Upsilon) \right\rangle = ieG_{\gamma}(q^2) \epsilon_{\mu\nu\alpha\beta}\epsilon_{\Upsilon}^\nu p^\alpha p^{\prime\beta},\nn
	T_\mu^X &\equiv& \left\langle \eta_b(p^\prime)|J_\mu^{X}|\Upsilon(p,\epsilon_\Upsilon) \right\rangle = ieG_{X}(q^2) \epsilon_{\mu\nu\alpha\beta}\epsilon_\Upsilon^\nu p^\alpha p^{\prime\beta}.
\end{eqnarray}

The full amplitude squared is written as
\begin{equation}
	|\mathcal{M}|^2 =|\mathcal{M}^\gamma+\mathcal{M}^X| = |\mathcal{M}^\gamma|^2 + |\mathcal{M}^X|^2 + 2\Re(\mathcal{M}^\gamma \mathcal{M}^{X*}).
\end{equation}
Note that the interference of the amplitudes is negligible because the total width of the $X17$ boson is very narrow~\cite{Castro:2021gdf}. Therefore, one can assume $ \Gamma \approx \Gamma^\gamma + \Gamma^X$ for simplicity.
The differential decay rates induced by photon and $X17$ boson are given by
\begin{eqnarray}
	\label{eq:dGamG}
	\frac{d\Gamma^\gamma}{dq^2} &=& \frac{\alpha^2_{\textrm{em}}}{72\pi m_{\Upsilon}^3} G^2_{\gamma}(q^2)\frac{1}{q^2}\left(1+\frac{2m_e^2}{q^2}\right)\sqrt{1-\frac{4m_e^2}{q^2}}\lambda^{3/2}(m_{\Upsilon}^2,m_{\eta_b}^2,q^2),\\
	\frac{d\Gamma^X}{dq^2} &=& \frac{\alpha^2_{\textrm{em}}\varepsilon_e^2}{72\pi m_{\Upsilon}^3} G^2_{X}(q^2)\frac{q^2}{(q^2-m_X^2)+m_X^2\Gamma_X^2}\left(1+\frac{2m_e^2}{q^2}\right)\sqrt{1-\frac{4m_e^2}{q^2}}\nn
	&\times&\lambda^{3/2}(m_{\Upsilon}^2,m_{\eta_b}^2,q^2),
	\label{eq:dGamX}
\end{eqnarray}
where $\lambda(x,y,z) \equiv x^2+y^2+z^2-2(xy+yz+zx)$ is 
the K{\"a}ll{\'e}n function.  
\begin{figure}[h]
	\centering	
	\includegraphics[width=0.45\textwidth]{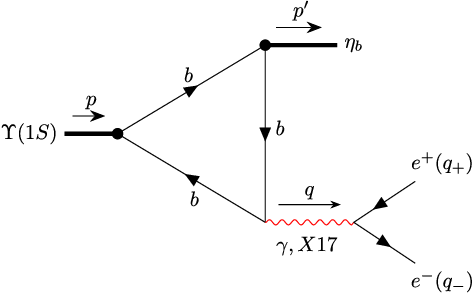}
	\vspace*{-3mm}	
	\caption{\label{fig:Dalitz}
		Feynman diagram for Dalitz decay $\Upsilon(1S)\to \eta_b e^+ e^-$.}
\end{figure}

The key element in theoretical calculations of $\Upsilon(1S)\to \eta_b\gamma$ and  $\Upsilon(1S)\to \eta_b e^+ e^-$ is the invariant form factors describing the $\Upsilon(1S)\to \eta_b$ hadronic transition. These are functions of momentum transfer between ititial and final hadrons that encapsulate the complex internal structure and dynamics of hadrons during their transitions and therefore determine the transition amplitudes. In general, the calculation of hadronic form factors requires nonperturbative methods such as lattice QCD (LQCD), QCD sum rules (QCDSR), and phenomenological quark models. In this study, we use the CCQM for the form factor evaluation. The CCQM is a phenomenological quark model built upon an interacting Lagrangian between a hadron and its constituent quarks. The model has been developed by our group and applied to various decays of hadrons (see, e.g., Refs.~\cite{Branz:2009cd, Ivanov:2011aa, Groote:2021ayy, Tran:2024phq}). We refer the reader to our recent studies~\cite{Tran:2025fhb,Tran:2023hrn} for a more detailed description of radiative and Dalitz decays within the CCQM. In this brief contribution, we only summarize the theoretical formalism presented in Refs.~\cite{Branz:2009cd,Tran:2025fhb,Tran:2023hrn} and focus more on the new results.

In the CCQM, the transition form factors are calculated based on the triangle Feynman diagram in Fig.~\ref{fig:Dalitz}. The hadronic amplitude $T_\mu^\gamma$ is written as
\begin{eqnarray}
	\left\langle \eta_b(p^\prime)|J_\mu^{\textrm{em}}|\Upsilon(p,\epsilon_\Upsilon) \right\rangle &=& (-3i)e g_{\Upsilon} g_{\eta_b} \epsilon_{\Upsilon}^\nu(p) (2e_b \mathcal{M}_{\mu\nu}),\\
	\mathcal{M}_{\mu\nu} &=& \int \frac{dk}{(2\pi)^4i}\widetilde{\Phi}_{\Upsilon}\big[-(k-\frac12 p)^2\big]\widetilde{\Phi}_{\eta_b}\big[-(k-\frac12 p^\prime)^2\big]\nonumber\\
	&&\times\mathrm{tr}\big[S_b(k)\gamma_\nu S_b(k-p)\gamma_\mu S_b(k-p^\prime)\gamma^5\big].
\end{eqnarray}
Here, $g_{\Upsilon}$ and $g_{\eta_b}$ are the meson-quark coupling constants obtained from the compositeness 
condition~\cite{Salam:1962ap,Weinberg:1962hj}
$Z_M = 1 - \Pi^\prime_M(m^2_M) = 0$,
where $Z_M$ is the meson's wave function renormalization constant and $\Pi'_M$ is the derivative of the meson's mass function. The vertex functions $\widetilde{\Phi}_{\Upsilon,\eta_b}$ effectively describes the meson's size and are assumed to be Gaussian for simplicity
\begin{equation}
	\widetilde{\Phi}_M(-p^2)=\exp(p^2/\Lambda^2_M),
	\label{eq:vertexf}
\end{equation}
where $\Lambda_M$ is the size parameter of the meson $M$. Note that the specific mathematical form for $\widetilde{\Phi}_M(-p^2)$ is not critical, as long as it decreases rapidly enough at high momentum values (in the Euclidean space) to ensure the ultraviolet finite of Feynman diagrams. The loop integration is performed by using the Fock-Schwinger representation for the quark propagator
\begin{equation}
	S_{b} (k) = (m_{b} + \not\! k)\int\limits_0^\infty \!\!d\alpha_i\,e^{-\alpha_i (m_{b}^2-k^2)}.
	\label{eq:Fock}
\end{equation}
Finally, the expression for the form factor $G_{\gamma(X)}(q^2)$ in the CCQM is given by
\begin{eqnarray}
	\label{eq:ff}
	G_{\gamma(X)}(q^2) &=& e_b(\varepsilon_b) g_{\Upsilon} g_{\eta_b} \frac{N_c}{4\pi^2}\int\limits_0^{1/{\lambda^2}} \!\! 
	\frac{dt\,t^2}{(s+t)^2} \int\limits\!\!d\alpha^3 \delta\big(1-\sum\limits_{i=1}^3\alpha_i \big)m_b \exp\left(-tz_0 + \frac{st}{s+t}z_1\right),\nonumber
	\\
	z_0 &=&  m_b^2-\alpha_1\alpha_2 m_{\Upsilon}^2-\alpha_2\alpha_3 m_{\eta_b}^2,
	\nn 
	z_1 &=& m_{\Upsilon}^2\left(\alpha_1-\frac{1}{2}\frac{s_{\Upsilon}}{s}\right)\left(\frac12-\alpha_2\right)+m_{\eta_b}^2\left(\alpha_2-\frac12\right)\left(\alpha_1+\alpha_2-\frac{s_{\Upsilon}}{s}-\frac12\frac{s_{\eta_b}}{s}\right),
	\nn
	s &=& s_{\Upsilon}+s_{\eta_b} , \qquad s_{\Upsilon{(\eta_b)}}=1/\Lambda^2_{\Upsilon{(\eta_b)}}.
\end{eqnarray}	
Here, we have introduced an infrared cutoff parameter $\lambda$ to avoid any possible thresholds in the Feynman diagram. This parameter is taken to be universal in our model. It effectively guarantees the confinement of quarks within hadrons~\cite{Branz:2009cd}.  

\section{Numerical results}
\label{sec:result}
Free parameters in the CCQM include the constituent quark masses $m_q$, the hadron size parameters $\Lambda_H$, and the cutoff parameter $\lambda$. Their values are determined by fitting the model's predictions to experimental data. In this work, the required parameters read $m_b = 5.05$~GeV, $\Lambda_{\Upsilon} = \Lambda_{\eta_b} = 4.03$~GeV, and $\lambda = 0.181$~GeV~\cite{Dubnicka:2024geu}. Other model-independent inputs are taken from the Particle Data Group~\cite{ParticleDataGroup:2024cfk}. 
The theoretical reliability of the calculated branching fractions depends on the precision of the underlying CCQM parameters. These parameters are determined through a $\chi^2$ minimization fit to experimental data for leptonic and electromagnetic decay constants, with the resulting parameter uncertainties typically restricted to the 2--5\% range. Rather than employing a complex error propagation analysis, we estimate the theoretical uncertainty of our predictions by evaluating the model's historical performance against experimental benchmarks. In previous applications of the CCQM to similar hadronic transitions, predicted quantities have consistently shown a deviation of approximately 5--10\% from measured values. Accounting for the cumulative effect of these variations on the transition form factors and the subsequent decay widths, we assign a conservative theoretical uncertainty of approximately 10\% to the decay widths and branching fractions presented in this work. 

Using constraints from ATOMKI anomalies ~\cite{Krasznahorkay:2015iga,Krasznahorkay:2021joi,Krasznahorkay:2022pxs} and from the NA48/2 experiment~\cite{NA482:2015wmo}, Denton and Gehrlein~\cite{Denton:2023gat} obtained the allowed regions for the quark--$X17$ coupling constants $|\varepsilon_u|\approx (0.5 - 0.9)\times 10^{-3}$ and $|\varepsilon_d|\approx (2.5 - 2.9)\times 10^{-3}$ with $\varepsilon_u \varepsilon_d <0$. If isospin mixing and breaking effects are taken into account, the favored values read $\varepsilon_u = \pm 9.0\times 10^{-4}$ and $\varepsilon_d = \mp 2.5\times 10^{-3}$. Assuming universal couplings between the $X17$ boson with quarks of the same type (up-type or down-type), we take $\varepsilon_b = \varepsilon_d = - 2.5\times 10^{-3}$ in this study.

Regarding the choice of the $b$-quark coupling, we assume $\varepsilon_b = \varepsilon_d = -2.5 \times 10^{-3}$. While the couplings of the $X17$ boson to different quark generations are \emph{a priori} independent, experimental constraints are currently concentrated on the first generation ($\varepsilon_u, \varepsilon_d$) via the ATOMKI nuclear transitions~\cite{Krasznahorkay:2015iga,Krasznahorkay:2021joi,Krasznahorkay:2022pxs} and the NA48/2 null result for $\pi^0 \to X \gamma$~\cite{NA482:2015wmo}. In the absence of direct data for the third generation, flavor universality within the down-type and up-type sectors, where $\varepsilon_b = \varepsilon_s = \varepsilon_d$ and $\varepsilon_c = \varepsilon_u$, provides a motivated and conservative starting point often used in the literature to facilitate theoretical comparisons. This assumption can be tested as more data from the heavier sectors becomes available. For instance, the BESIII Collaboration recently reported an upper limit on the charm coupling, $|\varepsilon_c| < 1.2 \times 10^{-2}$~\cite{BESIII:2025otp}, from charmonium transitions. While this limit is an order of magnitude larger than the universality-based estimate, it underscores the current lack of sensitivity compared to the first generation. Furthermore, alternative coupling assignments, assuming non-universal values for the first and second generations, can be derived from fits to other vector meson Dalitz decays~\cite{Lee:2025lwv}, yet the specific determination of $\varepsilon_b$ remains impossible without experimental data for $B$ meson or bottomonium Dalitz decays. Until such measurements are available, our adoption of $\varepsilon_b = \varepsilon_d$ serves as a standard baseline for assessing the sensitivity of $\Upsilon(1S)$ to NP contributions.

\begin{figure}[htbp]
	\centering
	\begin{tabular}{cc}
		\includegraphics[width=0.5\textwidth]{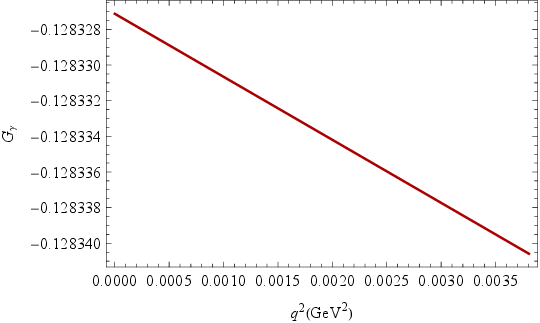}&
		\includegraphics[width=0.5\textwidth]{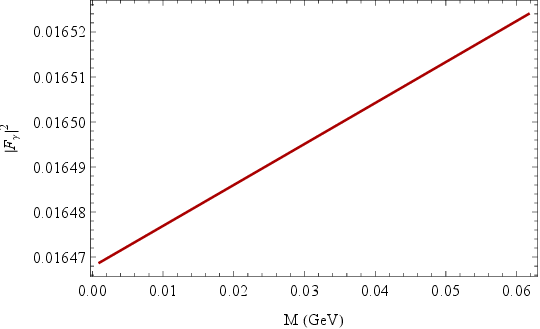}
	\end{tabular}
	\vspace*{-3mm}
	\caption{$\Upsilon(1S)\to \eta_b e^+ e^-$ form factors in the CCQM.}
	\label{fig:FF}
\end{figure}

The calculated form factors can be parametrized by a single-pole function of the form
\begin{equation}
	\label{eq:monopole}
	G_{\gamma (X)}(q^2)=\frac{G_{\gamma (X)}(0)}{1-aq^2}.
\end{equation}
The parameters read $G_{\gamma}(0) = -0.128$, $G_{X}(0) = -0.962\times 10^{-3}$, and $a = 0.028$. Since there is only one quark flavor in the loop, the form factors are simply related by $G_{\gamma}(q^2)/G_{X}(q^2) = e_b/\varepsilon_b$.
In the literature, the decay width distribution of the Dalitz decay within the SM is often written as~\cite{Gu:2019qwo, Tan:2021clg}
\begin{eqnarray}
	\frac{d\Gamma(\Upsilon\to \eta_b
		e^+e^-)}{dq^2} &=& \frac{\alpha_{\textrm{em}}}{3\pi q^2}\sqrt{1-\frac{4m^2_e}{q^2}}\left(1+\frac{2m^2_e}{q^2}\right)\left[1-\frac{q^2}{(m_{\Upsilon}-m_{\eta_b})^2}\right]^{\frac32}\left[1-\frac{q^2}{(m_{\Upsilon}+m_{\eta_b})^2}\right]^{\frac32}\nonumber\\
	&\times& \Gamma(\Upsilon\to \eta_b\gamma)	|F_\gamma(q^2)|^2,\nonumber\\
	&\equiv& [\textrm{QED}(q^2)]\times |F_\gamma(q^2)|^2,
	\label{eq:dGamSM}
\end{eqnarray}
where $	F_\gamma(q^2) \equiv G_\gamma(q^2)/G_\gamma(0)$ is the normalized form factor. The value $G_\gamma(0)$ is the radiative decay constant of the corresponding radiative decay $\Upsilon(1S)\to \eta_b\gamma$. Experimentally, the squared form factor $|F_\gamma(M)|^2$, where $M \equiv \sqrt{q^2}$ is the dilepton mass, can be measured by comparing the invariant mass distribution of the lepton pairs produced in the Dalitz decays to the prediction from QED for a point-like interaction~\cite{NA60:2009una}. We therefore plot both $G_{\gamma}(q^2)$ and  $|F_\gamma(M)|^2$ in Fig.~\ref{fig:FF}. There is no need to plot $G_{X}(q^2)$ since its behavior is exactly similar to $G_{\gamma}(q^2)$. 

The width of the Dalitz decay is calculated by integrating the Eqs.~(\ref{eq:dGamG}) and~(\ref{eq:dGamX}) over the physical range for the momentum transfer  $4m^2_e \leq q^2 \leq (m_{\Upsilon}-m_{\eta_b})^2$. Within the SM, we predict
\begin{eqnarray}
	\Gamma(\Upsilon(1S)\to\eta_b e^+ e^-) &=& 4.8(5)\times 10^{-2}\, \textrm{eV},\nonumber\\
	\mathcal{B}(\Upsilon(1S)\to\eta_b e^+ e^-) &=& 8.9(9)\times 10^{-7}.
\end{eqnarray}
\begin{figure}[htbp]
	\centering	
	\includegraphics[width=0.4\textwidth]{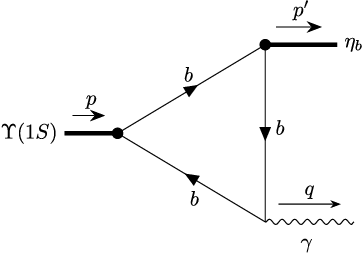}	
	\vspace*{-3mm}
	\caption{\label{fig:Rad}
		Feynman diagram for radiative decay $\Upsilon(1S)\to \eta_b \gamma$.}
\end{figure}
This small decay width is experimentally challenging. Instead of directly measuring the width, experiments such as CLEO~\cite{CLEO:2011mla} and BESIII~\cite{BESIII:2021vyq} usually measure the ratio of the Dalitz decay with respective to the corresponding radiative decay, defined as
\begin{equation}
	R_{ee} \equiv \frac{\Gamma(\Upsilon(1S)\to \eta_b e^+e^-)}{\Gamma(\Upsilon(1S)\to \eta_b\gamma)}.
\end{equation}

Within the CCQM, the radiative decay $\Upsilon(1S)\to \eta_b\gamma$ is described by the Feynman diagram in Fig.~\ref{fig:Rad}. The decay width is given by~\cite{Tran:2023hrn,Issadykov:2025gkf}
\begin{equation}
	\label{eq:rad-Gam}
	\Gamma(\Upsilon(1S)\to \eta_b \gamma) = \frac{\alpha_{\textrm{em}}}{24}  m_{\Upsilon}^3\left(1-
	\frac{m_{\eta_b}^2}{m_{\Upsilon}^2}\right)^3G_{\gamma}^2(0).
\end{equation}
Our predictions for the radiative decay read
\begin{eqnarray}
	\label{eq:RadGam}
	\Gamma(\Upsilon(1S)\to\eta_b \gamma) &=& 9.3(9)\, \textrm{eV},\nonumber\\
	\mathcal{B}(\Upsilon(1S)\to\eta_b \gamma) &=& 1.7(2)\times 10^{-4}.
\end{eqnarray}
Our predictions for $\Upsilon(1S)\to\eta_b \gamma$ in Eq.~(\ref{eq:RadGam}) are in good agreement with those in other theoretical approaches, such as the relativistic quark model ($\Gamma = 5.8~\textrm{eV}$, $\mathcal{B}=1.1\times 10^{-4}$)~\cite{Ebert:2002pp}, the light-front quark model (LFQM) ($\mathcal{B}=1.94(41)\times 10^{-4}$)~\cite{Ke:2010vn}, nonrelativistic effective QCD ($\Gamma = 3.6(2.9)~\textrm{eV}$, $\mathcal{B}=6.8(5.5)\times 10^{-4}$)~\cite{Brambilla:2005zw}, and the covariant Blankenbecler-Sugar equation ($\Gamma = 7.7~\textrm{eV}$)~\cite{Lahde:2002wj} (see also Refs.~\cite{Deng:2016ktl, Pandya:2014qma}). Notably, our predictions are very close to the values $\Gamma = 8.953~\textrm{eV}$ obtained by the Quarkonium Working Group based on nonrelativistic QCD (NRQCD)~\cite{QuarkoniumWorkingGroup:2004kpm}, $\Gamma = 9.29(91)~\textrm{eV}$ within the LFQM~\cite{Ridwan:2024ngc}, and  $\Gamma = 10~\textrm{eV}$, $\mathcal{B}=1.9\times 10^{-4}$ within the relativized quark model~\cite{Godfrey:2015dia}. Note that some studies predict larger values for the width/branching of $\Upsilon(1S)\to\eta_b \gamma$, such as $\Gamma = 15.18(51)~\textrm{eV}$ based on the potential NRQCD approach~\cite{Pineda:2013lta} (see also Refs.~\cite{Choi:2007se, Negash:2017rqt}). 
\begin{figure}[htbp]
	\centering	
	\includegraphics[width=0.65\textwidth]{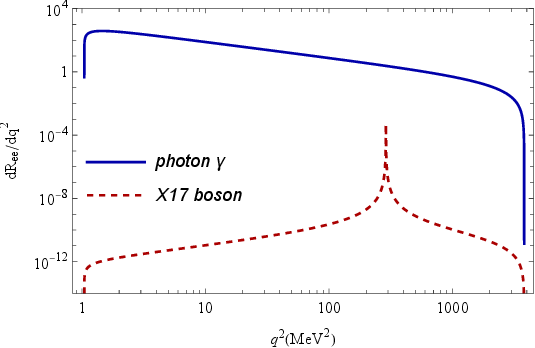}	
	\vspace*{-3mm}
	\caption{\label{fig:dRee}
		Differential decay rate for $\Upsilon(1S)\to \eta_b e^+e^-$. The solid/dashed line represents the contribution from $\gamma$/$X17$.}
\end{figure}

Finally, in Fig.~\ref{fig:dRee} we present the differential decay rates $dR^\gamma_{ee}/dq^2$ and $dR^X_{ee}/dq^2$ in the full physical range of $q^2$.
For the ratio $R_{ee}$, we predict the contributions from photon $R_{ee}^\gamma$ (within the SM) and from the $X17$ boson $R_{ee}^X$ to be
\begin{eqnarray}
	R_{ee}^\gamma &=& 5.1(5)\times 10^{-3},\nonumber\\
	R_{ee}^X &=& 5.0(5)\times 10^{-5},\nonumber\\
	R_{ee}^{\textrm{tot}} &\approx&  R_{ee}^\gamma + R_{ee}^X = 5.2(5)\times 10^{-3}.
\end{eqnarray}
The $X17$ contribution ($R_{ee}^X$) is two orders of magnitude smaller than the SM value ($R_{ee}^\gamma$).  Comparing with our previous study~\cite{Tran:2025fhb}, we conclude that the sensitivity of the decay $\Upsilon(1S)\to \eta_b e^+e^-$ to $X17$ is (i) similar to those of $B^{*0}_{(s)}\to B^0_{(s)} e^+e^-$ ($R_{ee}^X\sim 10^{-2}R_{ee}^\gamma$); (ii) larger than those of $D^{*0}\to D^0 e^+e^-$, $B^{*+}\to B^+ e^+e^-$, and $J/\psi\to \eta_c e^+e^-$ ($R_{ee}^X\sim 10^{-3}R_{ee}^\gamma$); and (iii) smaller than those of $D^{*+}_{(s)}\to D^+_{(s)} e^+e^-$ ($R_{ee}^X\sim 10^{-1}R_{ee}^\gamma$). The Dalitz decay $\Upsilon(1S)\to \eta_b e^+e^-$ can therefore serve as a promissing probe for the $X17$ vector boson at $B$--factories such as Belle~II and LHCb. Note that in the charm sector, similar experimental search for dark photon in the Dalitz decay of charmonium $J/\psi\to \eta^\prime e^+e^-$ has been done by the BESIII Collaboration~\cite{BESIII:2018aao}.

\section{Summary}
\label{sec:sum}
This contribution is an extension of our recent study~\cite{Tran:2025fhb} regarding the use of Dalitz decays $V\to P e^+ e^-$ as probes for the ATOMKI $X17$ vector boson. In this contribution, we have investigated the radiative and Dalitz decays $\Upsilon(1S)\to \eta_b\gamma$ and $\Upsilon(1S)\to \eta_b e^+ e^-$ of the bottomonium $\Upsilon(1S)$ with special focus on the ATOMKI anomaly. We have applied the Covariant Confined Quark Model to evaluate the transition form factors. We have provided predictions for the radiative decay constant, the branching fractions, and the decay rates. The contribution of the hypothetical $X17$ vector boson to the decay rate has been calculated. Our theoretical estimates presented in this work can serve as a valuable reference for upcoming experimental investigations into these decay modes and provide a benchmark for comparison with other theoretical frameworks in hadronic calculation.   

\section*{Acknowledgments}
This  research  is  funded  by  Vietnam National Foundation for Science and Technology Development (NAFOSTED) 
under grant number 103.01-2021.09. C.~T.~Tran thanks the organizers at Phenikaa University for the invitation and warm hospitality during the Phenikaa International Physics Conference (PIPC-2025).

	

\begin{thebibliography}{99}
	\bibitem{Krasznahorkay:2015iga}
	Krasznahorkay~A~J, Csatl\'os~M, Csige~L \textit{et al.} 2016
	Phys. Rev. Lett. \textbf{116} 042501 
	[arXiv:1504.01527].
	
	\bibitem{Krasznahorkay:2019lgi}
	Krasznahorkay~A~J, Csatl\'os~M, Csige~L \textit{et al.}
	2019 Acta Phys. Polon. B \textbf{50} 675. 
	
	\bibitem{Krasznahorkay:2021joi}
	Krasznahorkay~A~J, Csatl\'os~M, Csige~L \textit{et al.}
	2021 Phys. Rev. C \textbf{104} 044003 
	[arXiv:2104.10075].
	
	\bibitem{Krasznahorkay:2022pxs}
	Krasznahorkay~A~J, Krasznahorkay~A, Begala~M \textit{et al.}
	2022 Phys. Rev. C \textbf{106} L061601 
	[arXiv:2209.10795].
	
	\bibitem{Rose:1949zz}
	Rose~M~E 1949
	Phys. Rev. \textbf{76} 678
	[Erratum: 1950 Phys. Rev. \textbf{78} 184(E)].
	
	\bibitem{Schluter:1981cjo}
	Schl\"uter~P, Soff~G and Greiner~W
	1981 Phys. Rept. \textbf{75}, 327. 
	
	\bibitem{NA64:2018lsq}
	Banerjee~D \textit{et al.} (NA64)
	2018 Phys. Rev. Lett. \textbf{120} 231802 
	[arXiv:1803.07748].
	
	\bibitem{NA64:2019auh}
	Banerjee~D \textit{et al.} (NA64)
	2020 Phys. Rev. D \textbf{101} 071101 
	[arXiv:1912.11389]. 
	
	\bibitem{NA64:2020xxh}
	Depero~E \textit{et al.} (NA64)
	2020 Eur. Phys. J. C \textbf{80} 1159 
	[arXiv:2009.02756]. 
	
	\bibitem{MEGII:2024urz}
	Afanaciev~K \textit{et al.} (MEG~II)
	2025 Eur. Phys. J. C \textbf{85} 763 
	[arXiv:2411.07994].
	
	\bibitem{Anh:2024req}
	Anh~T, Trong~T~D, Krasznahorkay~A~J \textit{et al.}
	2024 Universe \textbf{10} 168 
	[arXiv:2401.11676]. 
	
	\bibitem{PADME:2025dla}
	Bossi~F \textit{et al.} (PADME)
	2025 JHEP \textbf{11} 007 
	[arXiv:2505.24797].
	
	\bibitem{PADME:2025dvz}
	Bertelli~S \textit{et al.} (PADME)
	2025 JHEP \textbf{06} 040 
	[arXiv:2503.05650]. 
	
	\bibitem{Alves:2023ree}
	Alves~D~S~M, Barducci~D, Cavoto~G \textit{et al.}
	2023 Eur. Phys. J. C \textbf{83} 230. 
	
	\bibitem{Feng:2016jff}
	Feng J L, Fornal B, Galon I \textit{et al.}
	2016 Phys. Rev. Lett. \textbf{117} 071803 
	[arXiv:1604.07411].
	
	\bibitem{Feng:2016ysn}
	Feng J L, Fornal B, Galon I \textit{et al.}
	2017 Phys. Rev. D \textbf{95} 035017 
	[arXiv:1608.03591]. 
	
	\bibitem{Zhang:2020ukq}
	Zhang X and Miller G A
	2021 Phys. Lett. B \textbf{813} 136061 
	[arXiv:2008.11288]. 
	
	\bibitem{Pulice:2019xel}
	Puli\c{c}e B
	2021 Chin. J. Phys. \textbf{71} 506
	[arXiv:1911.10482]. 
	
	\bibitem{Kozaczuk:2016nma}
	Kozaczuk J, Morrissey D E and Stroberg S R
	2017 Phys. Rev. D \textbf{95} 115024 
	[arXiv:1612.01525].
	
	\bibitem{Barducci:2022lqd}
	Barducci D and Toni C
	2023 JHEP \textbf{02} 154
	[Erratum: 2023 JHEP \textbf{07} 168(E)]
	[arXiv:2212.06453]. 
	
	\bibitem{Seto:2020jal}
	Seto O and Shimomura T
	2021 JHEP \textbf{04} 025 
	[arXiv:2006.05497]. 
	
	\bibitem{Nomura:2020kcw}
	Nomura T and Sanyal P
	2021 JHEP \textbf{05} 232
	[arXiv:2010.04266]. 
	
	\bibitem{Ellwanger:2016wfe}
	Ellwanger U and Moretti S
	2016 JHEP \textbf{11} 039 
	[arXiv:1609.01669]. 
	
	\bibitem{Kirpichnikov:2020tcf}
	Kirpichnikov D V, Lyubovitskij V E and Zhevlakov A S
	2020 Phys. Rev. D \textbf{102} 095024 
	[arXiv:2002.07496].
	
	\bibitem{DiLuzio:2025ojt}
	Di Luzio L, Paradisi P and Selimovic N
	2025 Nucl. Phys. B \textbf{1021} 117177
	[arXiv:2504.14014].
	
	\bibitem{Barducci:2025hpg}
	Barducci D, Germani D, Nardecchia M
	\textit{et al.}
	2025 JHEP \textbf{04} 035 
	[arXiv:2501.05507]. 
	
	\bibitem{Feng:2020mbt}
	Feng J L, Tait T M P and Verhaaren C B
	2020 Phys. Rev. D \textbf{102} 036016 
	[arXiv:2006.01151]. 
	
	\bibitem{Fu:2011yy}
	Fu J, Li H B, Qin X and Yang M Z
	2012 Mod. Phys. Lett. A \textbf{27} 1250223 
	[arXiv:1111.4055].
	
	\bibitem{Castro:2021gdf}
	Castro G L and Quintero N
	2021 Phys. Rev. D \textbf{103} 093002
	[arXiv:2101.01865].
	
	\bibitem{Ban:2020uii}
	Ban K, Jho Y, Kwon Y \textit{et al.}
	2021 JHEP \textbf{04} 091 
	[arXiv:2012.04190]. 
	
	\bibitem{Lee:2025lwv}
	Lee F F, Uyen L T T and Lin G L
	2026 Eur. Phys. J. C \textbf{86} 367 
	[arXiv:2501.13530].
	
	\bibitem{Colangelo:2025yud}
	Colangelo P, De Fazio F and Pinto R
	2026 Phys. Rev. D \textbf{113} 073007
	[arXiv:2512.17672]. 
	
	
	\bibitem{BESIII:2021vyq}
	Ablikim M \textit{et al.} (BESIII)
	2021 Phys. Rev. D \textbf{104} 112012 
	[arXiv:2111.06598]. 
	
	\bibitem{Tran:2025fhb}
	Tran C T, Ivanov M A and Nguyen T A T
	2025 Chin. Phys. \textbf{49} 113105
	[arXiv:2506.23372].
	
	\bibitem{CLEO:2011mla}
	Cronin-Hennessy D \textit{et al.} (CLEO)
	2012 Phys. Rev. D \textbf{86} 072005
	[arXiv:1104.3265].
	
	\bibitem{Branz:2009cd}
	Branz~T, Faessler~A, Gutsche~T \textit{et al.} 2010
	Phys. Rev. D \textbf{81} 034010 
	[arXiv:0912.3710].
	
	\bibitem{Ivanov:2011aa} 
	Ivanov~M~A, K\"orner~J~G, Kovalenko~S~G, Santorelli~P \textit{et al.}
	2012 Phys.\ Rev.\ D {\bf 85} 034004
	[arXiv:1112.3536].
	
	\bibitem{Groote:2021ayy}
	Groote~S, Ivanov~M~A, K\"orner~J~G, Lyubovitskij~V~E \textit{et al.} 2021
	Phys. Rev. D \textbf{103} 093001 
	[arXiv:2102.12818].
	
	\bibitem{Tran:2024phq}
	Tran~C~T, Ivanov~M~A, Santorelli~P and Tran~H~C 2025
	Chin. Phys. C \textbf{49} 013111 
	[arXiv:2408.13776].
	
	\bibitem{Tran:2023hrn}
	Tran~C~T, Ivanov~M~A, Santorelli~P and Vo~Q~C 2024
	Chin. Phys. C \textbf{48} 023103
	[arXiv:2311.15248].
	
	\bibitem{Salam:1962ap}
	Salam A  
	1962 Nuovo Cim. \textbf{25} 224.
	
	\bibitem{Weinberg:1962hj}
	Weinberg S
	1963 Phys. Rev. \textbf{130} 776.
	
	\bibitem{Dubnicka:2024geu}
	Dubni{\v{c}}ka S, Dubni{\v{c}}kov{\'a} A Z, Ivanov M A, Liptaj A \textit{et al.}
	2024 Phys. Rev. D \textbf{110}  056030 
	[arXiv:2406.09763].
	
	\bibitem{ParticleDataGroup:2024cfk}
	Navas~S \textit{et al.} (Particle Data Group) 2024
	Phys. Rev. D \textbf{110} 030001.
	
	\bibitem{NA482:2015wmo}
	Batley J R \textit{et al.} (NA48/2)
	2015 Phys. Lett. B \textbf{746} 178
	[arXiv:1504.00607].
	
	\bibitem{Denton:2023gat}
	Denton P B and Gehrlein J
	2023 Phys. Rev. D \textbf{108} 015009 
	[arXiv:2304.09877]. 
	
	\bibitem{BESIII:2025otp}
	Ablikim M \textit{et al.} (BESIII)
	2026 Phys. Rev. D \textbf{113} 032009
	[arXiv:2510.16531].
	
	\bibitem{Gu:2019qwo}
	Gu L M, Li H B, Ma X X \textit{et al.} 
	2019 Phys. Rev. D \textbf{100} 016018 
	[arXiv:1904.06085]. 
	
	\bibitem{Tan:2021clg}
	Tan Y, Zhang Z and Zhou X
	2022 Int. J. Mod. Phys. A \textbf{37} 2250075 
	[arXiv:2111.04932]. 
	
	\bibitem{NA60:2009una}
	Arnaldi R \textit{et al.} (NA60)
	2009 Phys. Lett. B \textbf{677} 260
	[arXiv:0902.2547].
	
	\bibitem{Issadykov:2025gkf}
	Issadykov A, Ivanov M A, Nguyen N D K \textit{et al.}
	2025 Phys. Rev. D \textbf{112}  094041 
	[arXiv:2509.13657].
	
	\bibitem{Ebert:2002pp}
	Ebert D, Faustov R N and Galkin V O
	2003 Phys. Rev. D \textbf{67} 014027 
	[arXiv:hep-ph/0210381].
	
	\bibitem{Ke:2010vn}
	Ke H W, Li X Q, Wei Z T and Liu X
	2010 Phys. Rev. D \textbf{82} 034023
	[arXiv:1006.1091].
	
	\bibitem{Brambilla:2005zw}
	Brambilla N, Jia Y and Vairo A
	2006 Phys. Rev. D \textbf{73} 054005 
	[arXiv:hep-ph/0512369].
	
	\bibitem{Lahde:2002wj}
	Lahde T A
	2003 Nucl. Phys. A \textbf{714} 183
	[arXiv:hep-ph/0208110].
	
	\bibitem{Deng:2016ktl}
	Deng W J, Liu H, Gui L C and Zhong X H
	2017 Phys. Rev. D \textbf{95} 074002 
	[arXiv:1607.04696].
	
	\bibitem{Pandya:2014qma}
	Pandya J N, Soni N R, Devlani N and Rai A K
	2015 Chin. Phys. C \textbf{39} 123101
	[arXiv:1412.7249].
	
	\bibitem{QuarkoniumWorkingGroup:2004kpm}
	Brambilla N \textit{et al.} (Quarkonium Working Group)
	arXiv:hep-ph/0412158. 
	
	\bibitem{Ridwan:2024ngc}
	Ridwan M, Arifi A J and Mart T
	2025 Phys. Rev. D \textbf{111} 016011 
	[arXiv:2409.13172].
	
	\bibitem{Godfrey:2015dia}
	Godfrey S and Moats K
	2015 Phys. Rev. D \textbf{92} 054034 
	[arXiv:1507.00024].
	
	\bibitem{Pineda:2013lta}
	Pineda A and Segovia J
	2013 Phys. Rev. D \textbf{87}  074024 
	[arXiv:1302.3528].
	
	\bibitem{Choi:2007se}
	Choi H M
	2007 Phys. Rev. D \textbf{75} 073016 
	[arXiv:hep-ph/0701263].
	
	\bibitem{Negash:2017rqt}
	Negash H and Bhatnagar S
	2017 Adv. High Energy Phys. \textbf{2017} 7306825 
	[arXiv:1703.06082].
	
	\bibitem{BESIII:2018aao}
	Ablikim M \textit{et al.} (BESIII)
	2019 Phys. Rev. D \textbf{99} 012013 
	[arXiv:1809.00635].
\end{thebibliography}
\end{document}